\documentclass[english,utf8,biblatex]{lni}
\usepackage{booktabs}
\usepackage{comment}

\begin{document}

% \title[To divide or not to divide classes of gamma-ray sources?]{To divide or not to divide classes of gamma-ray sources?}
% Should we use this title instead?
\title[Multiclass classification of gamma-ray sources]
{Towards probabilistic multiclass classification of gamma-ray sources}

\author[Dmitry Malyshev \and Aakash Bhat]{%
Dmitry Malyshev\footnote{Erlangen Centre for Astroparticle Physics, Erwin-Rommel-Str. 1, Erlangen, Germany 
\email{dmitry.malyshev@fau.de}} \and
Aakash Bhat\footnote{Erlangen Centre for Astroparticle Physics, Erwin-Rommel-Str. 1, Erlangen, Germany 
\email{aakashbhat7@gmail.com}}
}
\startpage{1}
\editor{Editor et al.}
\booktitle{INFORMATIK 2022 Workshops}
\yearofpublication{2022}
\lnidoi{12.34567/provided-by-the-editors}
\maketitle

\begin{abstract}
%The abstract should contain between 70 and 150 words.

Machine learning algorithms have been used to determine probabilistic classifications of unassociated sources.
Often classification into two large classes, such as Galactic and extra-galactic, is considered.
However, there are many more physical classes of sources (23 classes in the latest Fermi-LAT 4FGL-DR3 catalog \cite{2022arXiv220111184F}).
In this note we subdivide one of the large classes into two subclasses in view of a more general multi-class classification of gamma-ray sources.
Each of the three large classes still encompasses several of the physical classes.
We compare the performance of classifications into two and three classes.
We calculate the receiver operating characteristic curves for two-class classification, where in case of three classes we sum the probabilities of the sub-classes in order to obtain the class probabilities for the two large classes.
We also compare precision, recall, and reliability diagrams in the two- and three-class cases.

\end{abstract}


\section{Introduction}

About one third of gamma-ray sources in Fermi Large Area Telescope (LAT) catalogs are unassociated \cite{2015ApJS..218...23A,2020ApJS..247...33A, 2022arXiv220111184F}.
Machine learning methods trained with associated sources have been used to determine most likely classes of unassociated sources or, more generally, the probabilities for unassociated sources to belong to different classes
\cite{2012ApJ...753...83A,2016ApJ...820....8S,2016ApJ...825...69M,2017A&A...602A..86L,2020MNRAS.492.5377L,2021MNRAS.507.4061F,2021RAA....21...15Z,2022A&A...660A..87B}.
One of the main challenges in this approach is that the number of sources is relatively small: 6658 sources in the Fourth Fermi LAT source catalog data release 3 (4FGL-DR3) \cite{2022arXiv220111184F}, while the number of classes is relatively large: 23 classes in 4FGL-DR3 excluding unassociated sources and sources with ``unknown'' class, i.e., sources within $10^\circ$ from the Galactic plane and associated to radio or X-ray sources of unknown nature \cite{2020ApJS..247...33A}.
In order to obtain sufficiently many sources for training and testing, the physical classes are combined in larger classes.
In most of the analyses two classes are used, such as extra-galactic and Galactic sources, or active galactic nuclei (AGNs) and pulsars.

Subdivision of large classes into subclasses (which we refer to as ``class subdivision'' in the rest of the paper for brevity) 
was considered in the past, e.g., for sub-classification of pulsars into young and millisecond ones \cite{2016ApJ...820....8S, 2020MNRAS.492.5377L} or for classification of Galactic sources into pulsars and other sources \cite{2021RAA....21...15Z}.
The procedure of \cite{2016ApJ...820....8S,2020MNRAS.492.5377L}
included classification of all sources into AGNs and pulsars first, followed by dividing likely pulsars into young and millisecond pulsars by a separate ML training step.
Apart from being a relatively complicated procedure,
which is difficult to generalize beyond 3 classes, there is also a conceptual problem that the results depend on the choice of the threshold for the separation of pulsars and AGNs.
Although such procedure makes sense for the search of the most likely candidates, the pulsar-like probabilities in sources classified as AGNs are not taken into account for the young vs millisecond pulsar classification, which prevents the interpretation of the final scores as probabilities and complicates the use of such catalogs for population studies including unassociated sources.
In \cite{2021RAA....21...15Z}, the 2-class classification included AGNs and pulsars, while the 3-class case included AGNs, pulsars, and other Galactic sources.
The results of \cite{2021RAA....21...15Z} for the 3-class classification with two different ML algorithms lacked consistency. For instance, the random forest (RF) method gave 244 source candidates in the ``other'' class for 20 input features, while artificial neural networks (NNs) gave no candidates for the same input data.
In \cite{2022A&A...660A..87B} we have revisited the question of the 3-class classification into AGNs, pulsars, and other sources, and have shown that it is possible to have a consistent classification of sources into the three classes by several ML algorithms.
Interestingly, we have also found that the reliability diagrams in the 3-class case and the systematic uncertainties related to the choice of the classification method were similar in the 3-class case compared to the 2-class case.
In this work we extend the consistency checks of the 3-class classification using the latest 4FGL-DR3 catalog \cite{2022arXiv220111184F}.
In particular, we address the following question: can the 3-class classification also be used for the 2-class classification, where the 2-class probabilities are obtained by summing the probabilities of sub-classes in the 3-class case? For instance, would receiver operating characteristic (ROC) curve, precision, and recall for the 2-class case reduced from the 3-class case be similar, better, or worse compared to these characteristics derived in the 2-class classification?%
\footnote{In this paper we use only associated sources to study the question of class subdivision. Application to probabilistic classification of unsassiciated sources will be reported in a follow up study.}
For example, if the performance of the 2- and 3-class classifications for 2-class characteristics is similar, then the 3-class classification has a clear advantage over the 2-class classification since the 2-class probabilities can be recovered from the 3-class classification and this procedure does not degrade the 2-class classification performance. On the other hand, the 3-class classification provides additional information about probabilities of subclasses.

In this work we start with two large classes (AGN-like and Galactic sources) and test the effects on classification performance of dividing
the Galactic sources into pulsars and other Galactic sources.
One of the challenges of subdividing the classes is due to dataset imbalance - there are about seven times more AGN-like sources than Galactic ones. 
Dividing the Galactic sources into two classes leads to further imbalance of the datasets with imbalance factors on the order of 13 to 1 for 
AGN-like vs pulsars or 15 to 1 for AGN-like vs other Galactic sources.
On the one hand the increased imbalance can lead to a decrease in performance in the two-class classification,
where the probabilities of the sub-classes of Galactic sources in the 3-class case are added to determine the overall probability for the source
to be Galactic.
On the other hand, the additional information contained in the sub-classes may lead to an improved two-class classification.
For the classification we use RF and NN (with a single hidden layer) algorithms.
We show that at least for the example considered in this paper, the 3-class classification does not lead to a decrease of performance compared to the 2-class
classification, even if the subdivision of classes leads to an increase of imbalance among the classes.

The paper is organized as follows. In Section \ref{sec:training} we provide details about the training of the RF and NN algorithms including a description of classes and
features. We also optimize the selection of meta-parameters by calculating the overall accuracy of predictions in the 2- and 3-class cases.
In Section \ref{sec:tests} we perform consistency checks of classification, when we sub-divide the Galactic class. In particular, we
\begin{enumerate}
\item 
Compare 2-class classification characteristics: (1) ROC curves and (2) precision and recall for Galactic sources
as a function of thresholds on probability to belong to the Galactic class in cases when we subdivide and do not subdivide the Galactic class;
\item
Calculate the precision and recall for the sub-classes of the Galactic class as a function of threshold on probability to belong to the corresponding sub-classes;
\item
Calculate the reliability diagrams in the 2- and 3-class cases.
\end{enumerate}
Section \ref{sec:conclusions} contains conclusions and discussion.

\section{Training}
\label{sec:training}

\subsection{Definition of classes in the 4FGL-DR3 catalog}

The 4FGL-DR3 catalog has 23 classes in addition to unassociated sources or sources with an unknown class \cite{2022arXiv220111184F}.
If a source is associated with a source at other frequencies and there is additional information, such as coincident flaring activity for AGNs or pulsed gamma-ray emission for pulsars, then the counterpart is said to be identified and the corresponding class is written in capital letters. Classes of sources based on positional associations only are written in lower-case letters.
In this work we will treat both types of associations on equal footing and use lower-case class names for both identified and associated sources.

As the first step, we divide the sources into two large classes:
\begin{enumerate}
\item AGNs -- 3814 sources (4FGL-DR3 class names: 'bll', 'fsrq', 'rdg', 'agn', 'ssrq', 'css', 'bcu', 'nlsy1', 'sey')%
\footnote{For the full description of the classes, please see \cite{2022arXiv220111184F}.};
\item Non AGN sources -- 553 sources (4FGL-DR3 class names: 'gc', 'psr', 'msp', 'pwn', 'snr', 'spp', 'glc', 'sfr', 'hmb', 'lmb', 'bin', 'nov', 'sbg', 'gal').
Since the majority of non-AGN sources are inside the Milky way galaxy,
we will denote this class as Galactic sources.
\end{enumerate}

At the next step we divide the Galactic sources into two sub-classes
\begin{enumerate}
\item Pulsars -- 292 sources (4FGL-DR3 class names: 'psr', 'msp');
\item Galactic non-pulsar sources -- 261 sources (4FGL-DR3 class names: 'gc', 'pwn', 'snr', 'spp', 'glc', 'sfr', 'hmb', 'lmb', 'bin', 'nov', 'sbg', 'gal').
\end{enumerate}

\subsection{Algorithms}

\begin{figure}
  \centering
  \includegraphics[width=.45\textwidth]{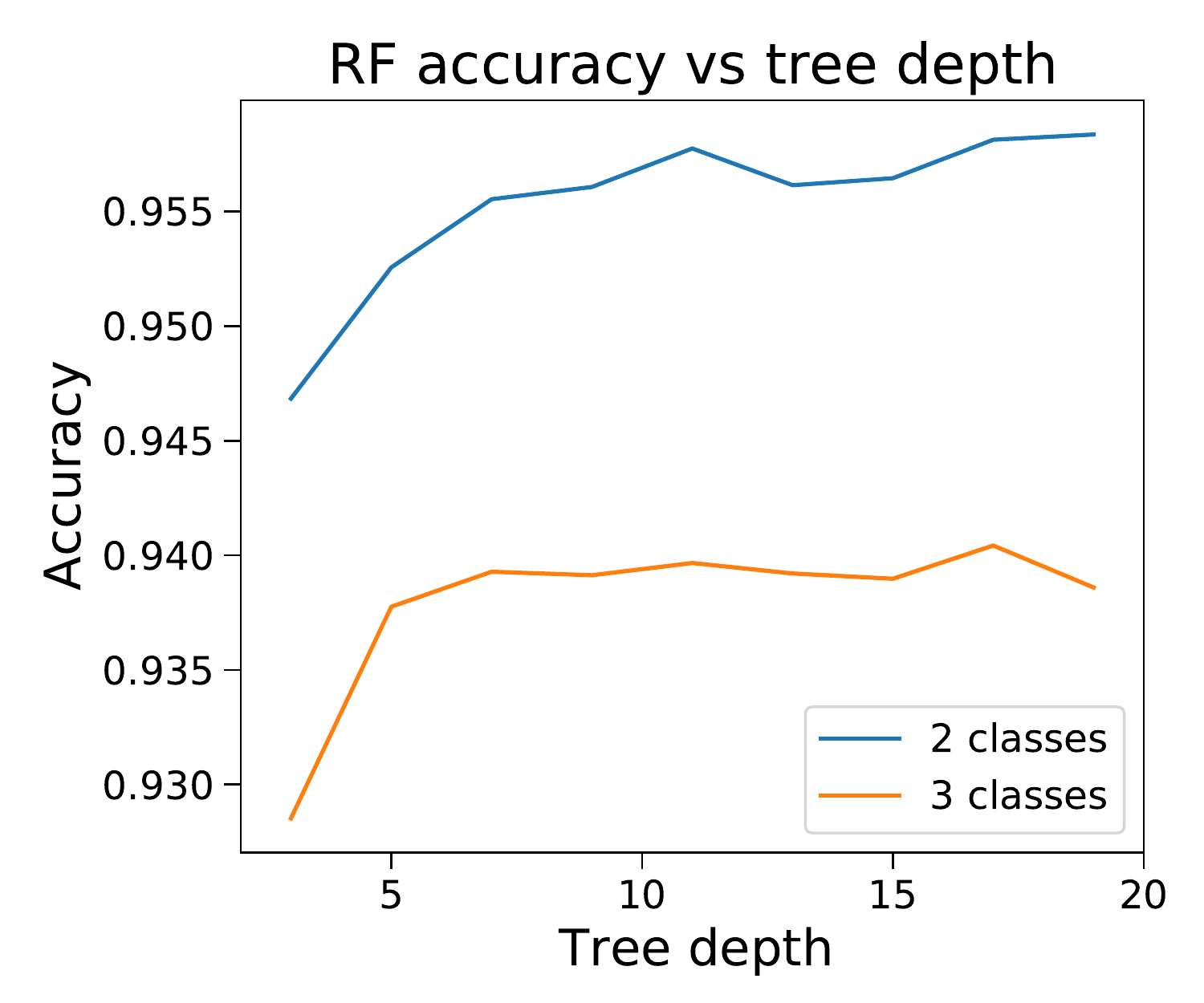}
  \includegraphics[width=.45\textwidth]{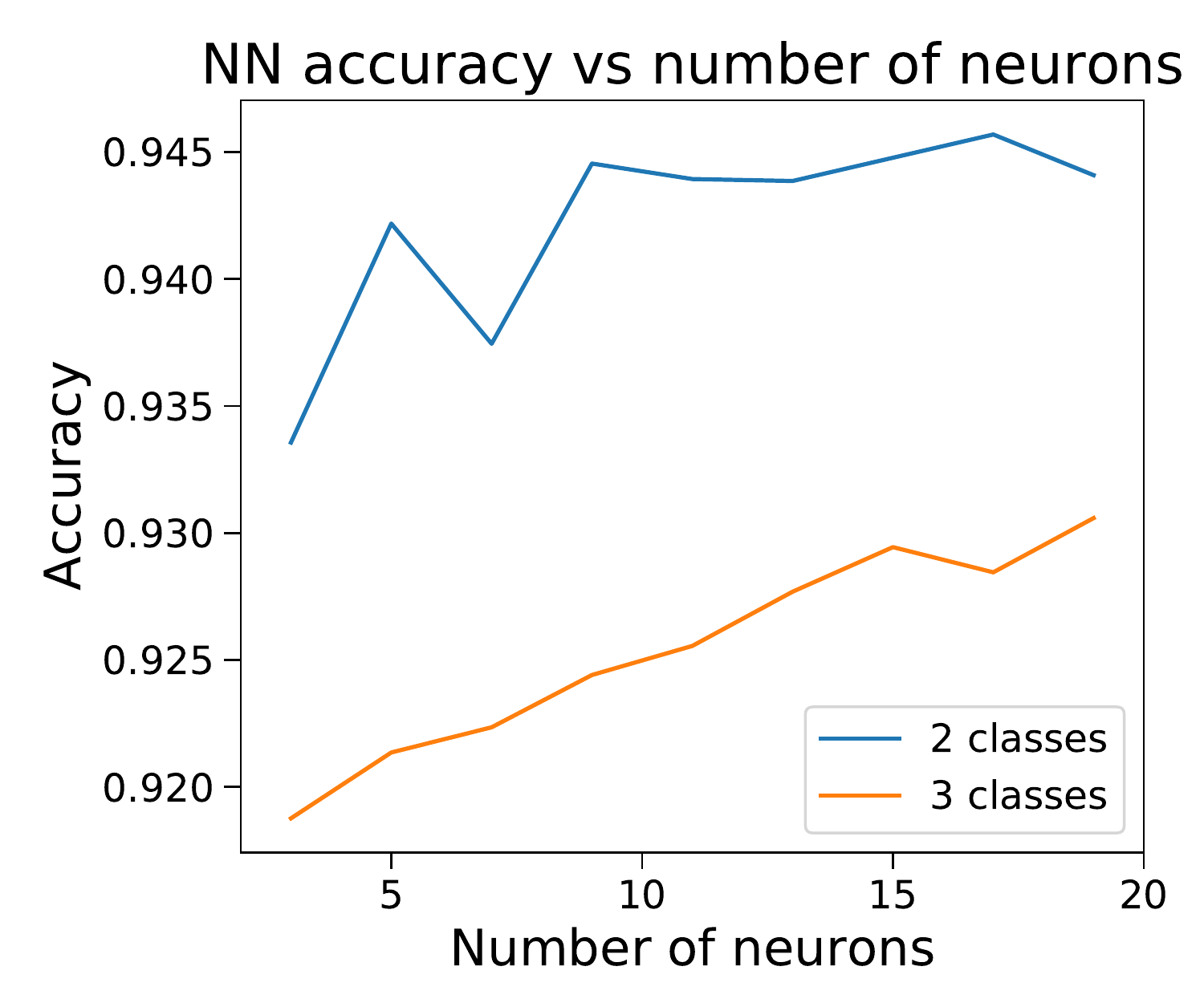}
  \caption{RF accuracy as a function of maximal tree depth (left) 
  and NN accuracy as a function of the number of neurons in the hidden layer (right).}
  \label{fig:depth}
\end{figure}

In this work we use the RF algorithm with 50 trees and a NN with a single hidden layer implemented in scikit-learn package \cite{scikit-learn}.
The NN has the ``tanh'' activation function at the hidden layer and ``softmax'' activation function for the output layer.
As input, we use the following 10 features from the 4FGL-DR3 catalog \cite{2022arXiv220111184F}
(some features are transformed in order to have a comparable range of values for the different features and in order to avoid discontinuity for the Galactic longitude): 
sin(GLAT), 
cos(GLON), 
sin(GLON), 
$\log_{10}$(Energy\_Flux100), 
$\log_{10}$(Unc\_Energy\_Flux100), 
$\log_{10}$(Signif\_Avg), 
LP\_beta,
LP\_SigCurv,
Variability\_Index,
and the index of the log parabola spectrum at 500 MeV%
\footnote{The definitions of the feature names can be found in the 4FGL-DR3 catalog \cite{2022arXiv220111184F}}.
The last feature is calculated as the derivative at 500 MeV:
\begin{equation}
    n = -\frac{d \log F(E)}{d \log E}
\end{equation}
where $F(E)$ is the log parabola spectrum of the source reported in the 4FGL-DR3 catalog.
These features characterize the main properties of gamma-ray sources, such as the position on the sky,
the shape of the energy spectrum, overall flux, and variability as a function of time.
We have also checked in an earlier analysis that these features have relatively small correlation among themselves, i.e., they are not redundant
\cite{2022A&A...660A..87B}.

We split the associated sources into 70\% training and 30\% testing samples.
We show the overall accuracy of classification in the 2- and 3-class cases as a function of maximal tree depth for the RF algorithm or
the number of neurons in the NN algorithm in \Cref{fig:depth}.
Each point on this plot is obtained by an average over 10 random splits into training and testing samples.
In the RF case, the accuracy saturates around the maximal depth of 10.
In the rest of the paper we use the RF algorithm with 50 trees of maximal depth 10.
For the NN algorithm, the accuracy saturates around 10 neurons in the hidden layer in the 2-class case or 15 neurons in the 3-class case.
In the following analysis we use 15 neurons in the hidden layer.

\section{Tests}
\label{sec:tests}

\subsection{ROC curves}

\begin{figure}
  \centering
  \includegraphics[width=.45\textwidth]{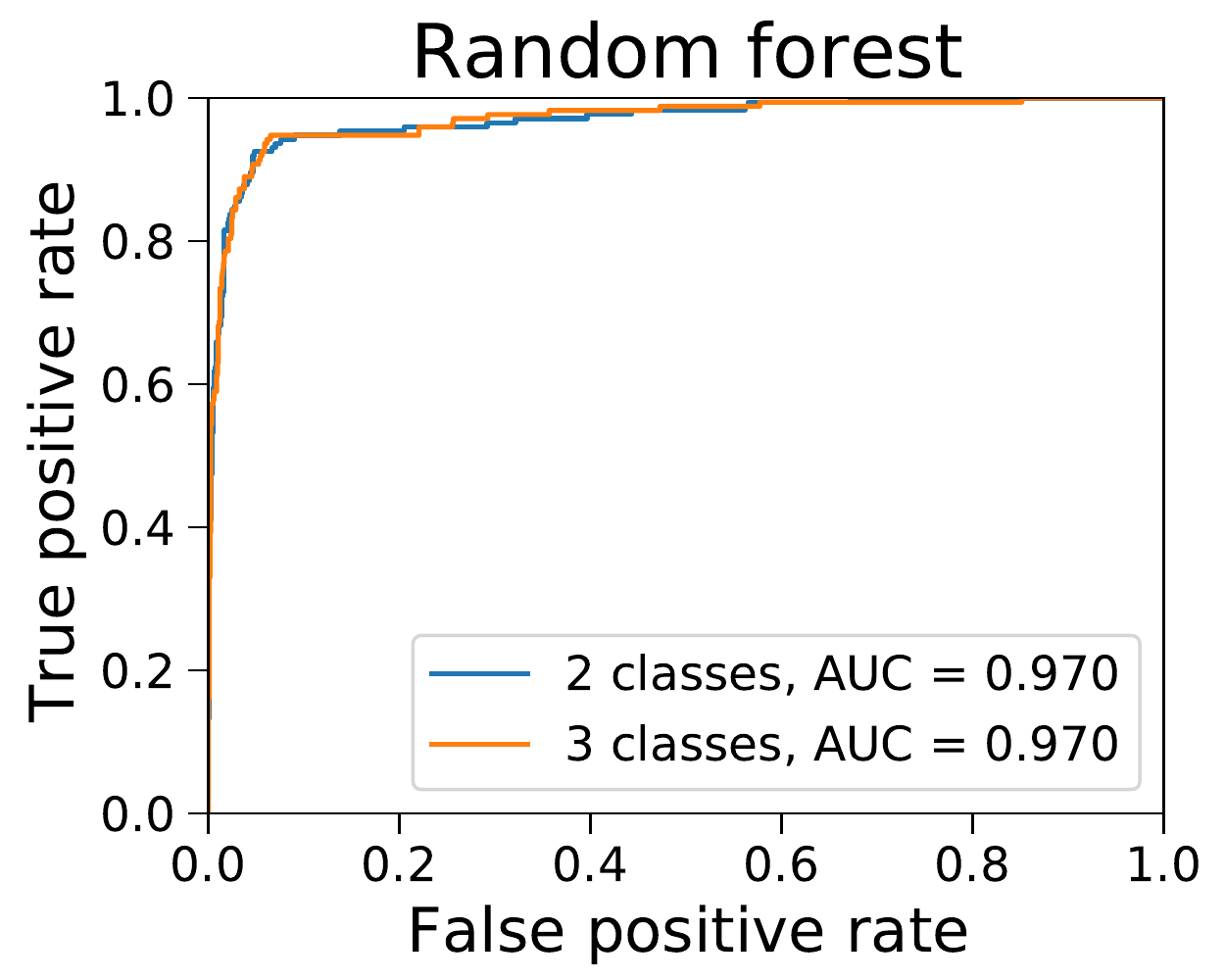}
  \includegraphics[width=.45\textwidth]{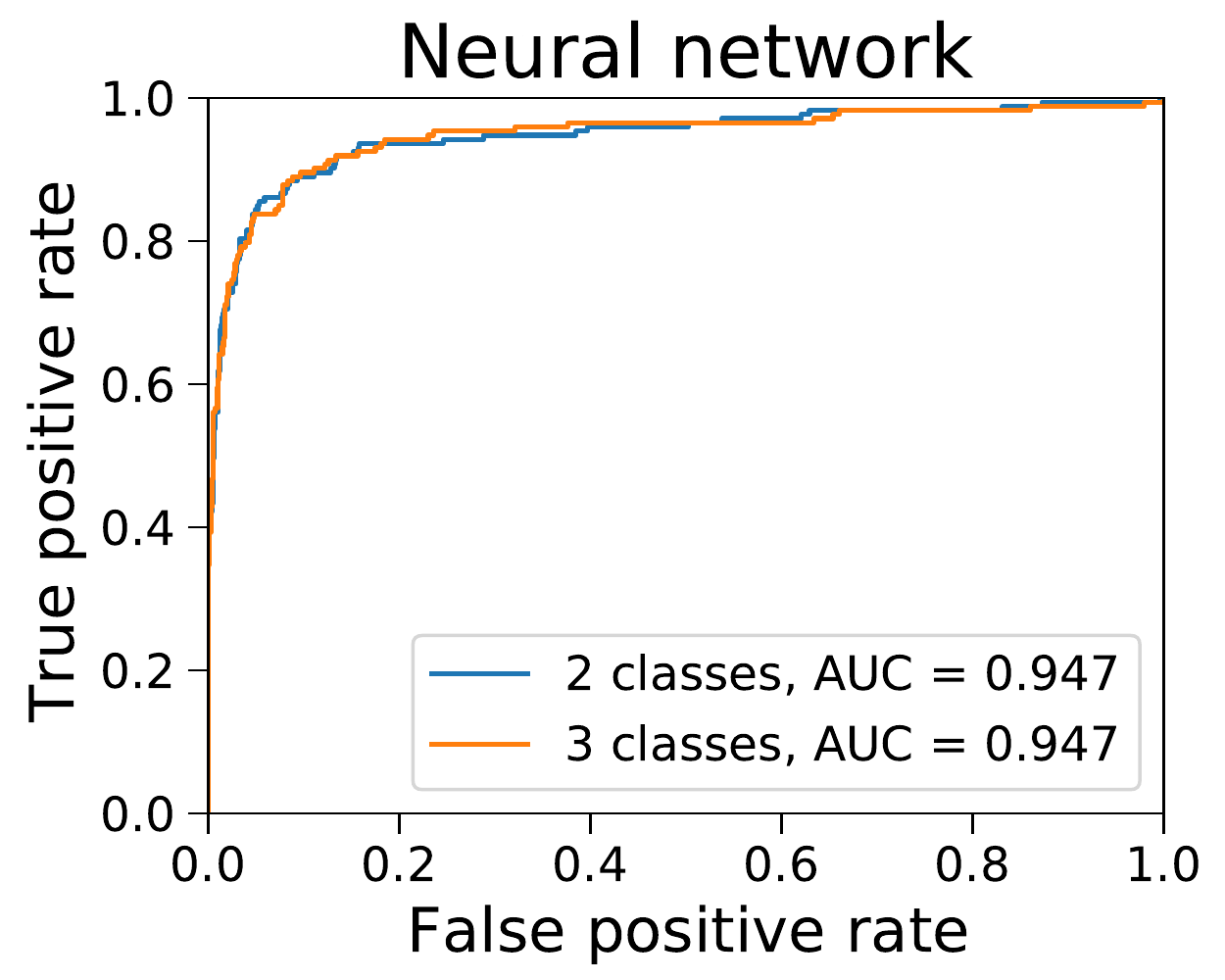}
  \caption{ROC curves for the Galactic class in the 2- and 3-class classification with RF (left) and NN (right) algorithms. 
  The 3 classes are obtained by dividing the Galactic class into two subclasses: pulsars and non-pulsar Galactic sources.
  In order to obtain the Galactic class probability in the 3-class case, we sum the probabilities of the two subclasses.
  }
  \label{fig:ROC}
\end{figure}

As a first test of the effect of dividing the Galactic class into two sub-classes,
we calculated the ROC curve in the 2- and 3-class classifications.
Here and in the following sections we use sources from the test sample for the comparison of the 2- and 3-class classifications.
In the 3-class case, we sum the probabilities for the two sub-classes of the Galactic class in order to determine the probability for the sources to belong to the Galactic class.
We find that there is essentially very little difference in the 2- or 3-class classifications with regards to the 2-class separation (\Cref{fig:ROC}).

\subsection{Precision and recall}

\begin{figure}
  \centering
  \includegraphics[width=0.45\textwidth]{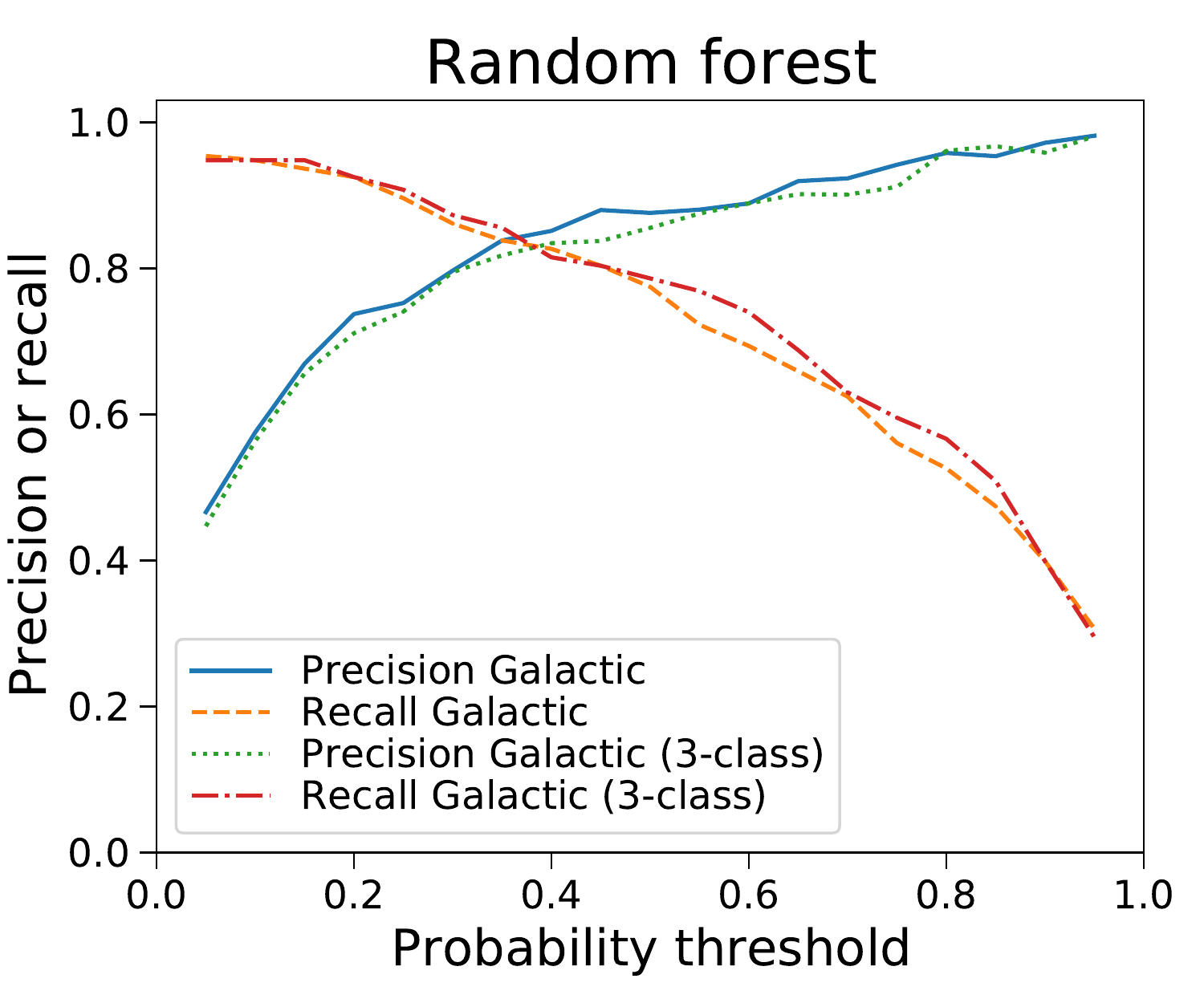}
  \includegraphics[width=0.45\textwidth]{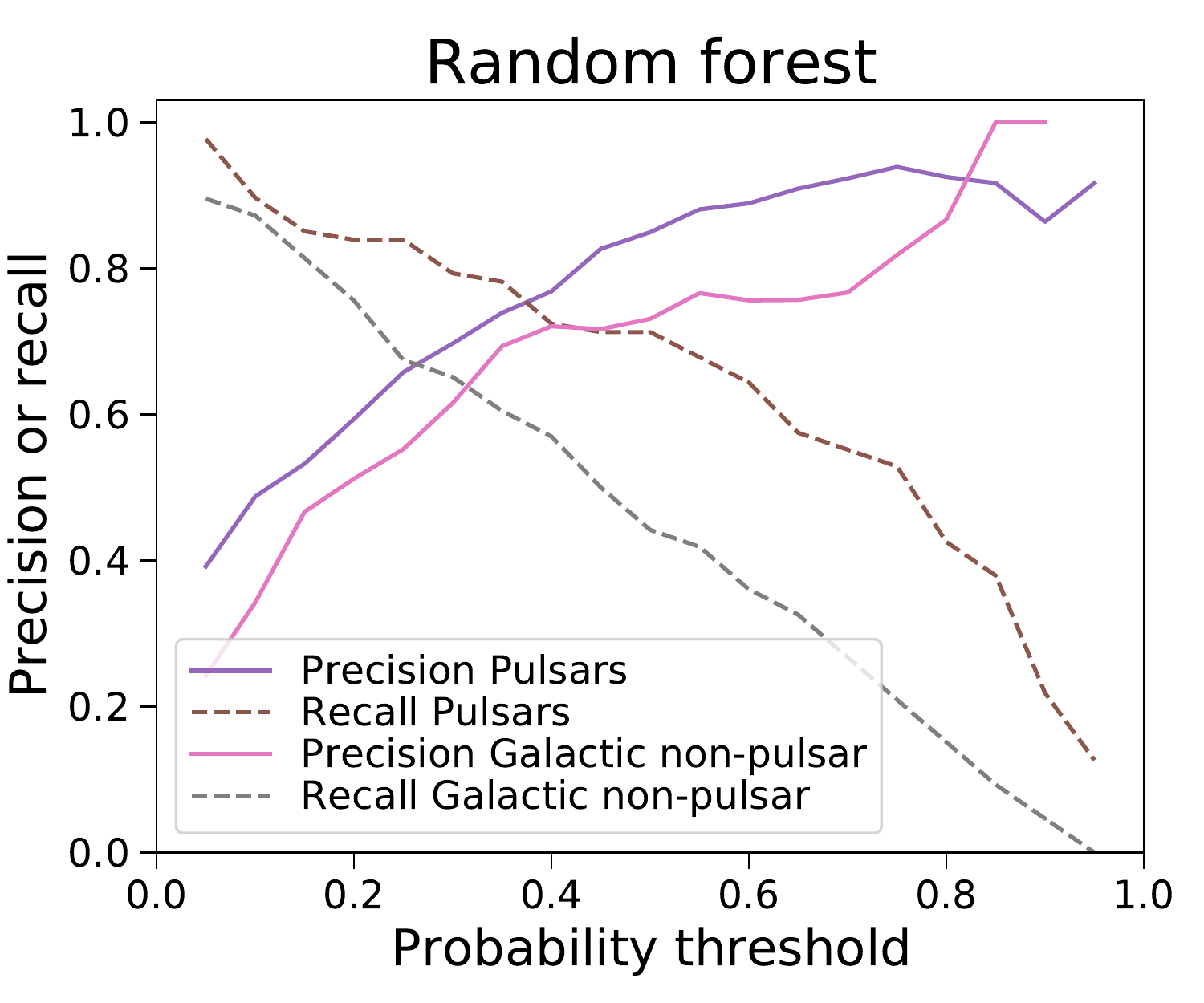}\\
  \includegraphics[width=0.45\textwidth]{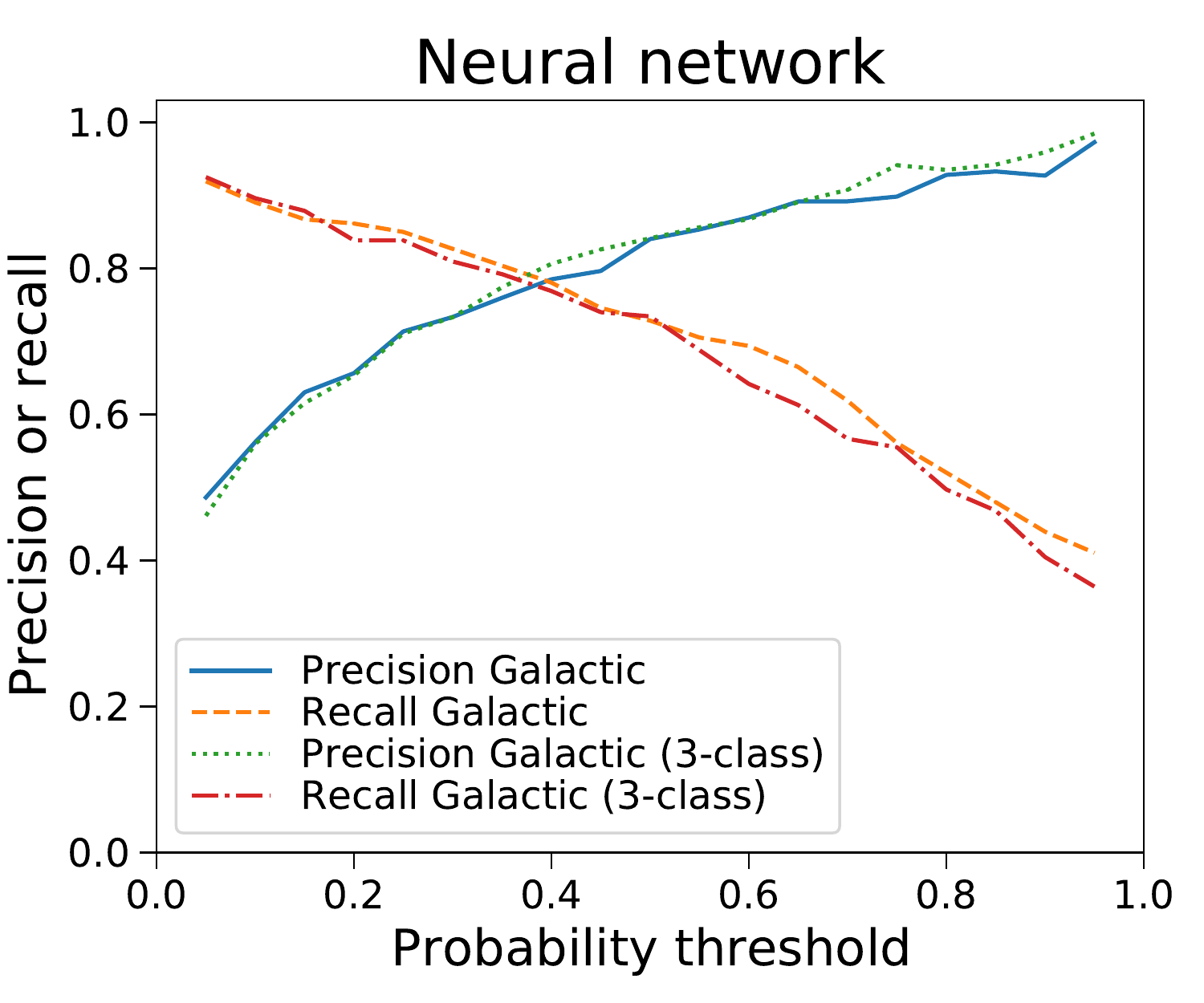}
  \includegraphics[width=0.45\textwidth]{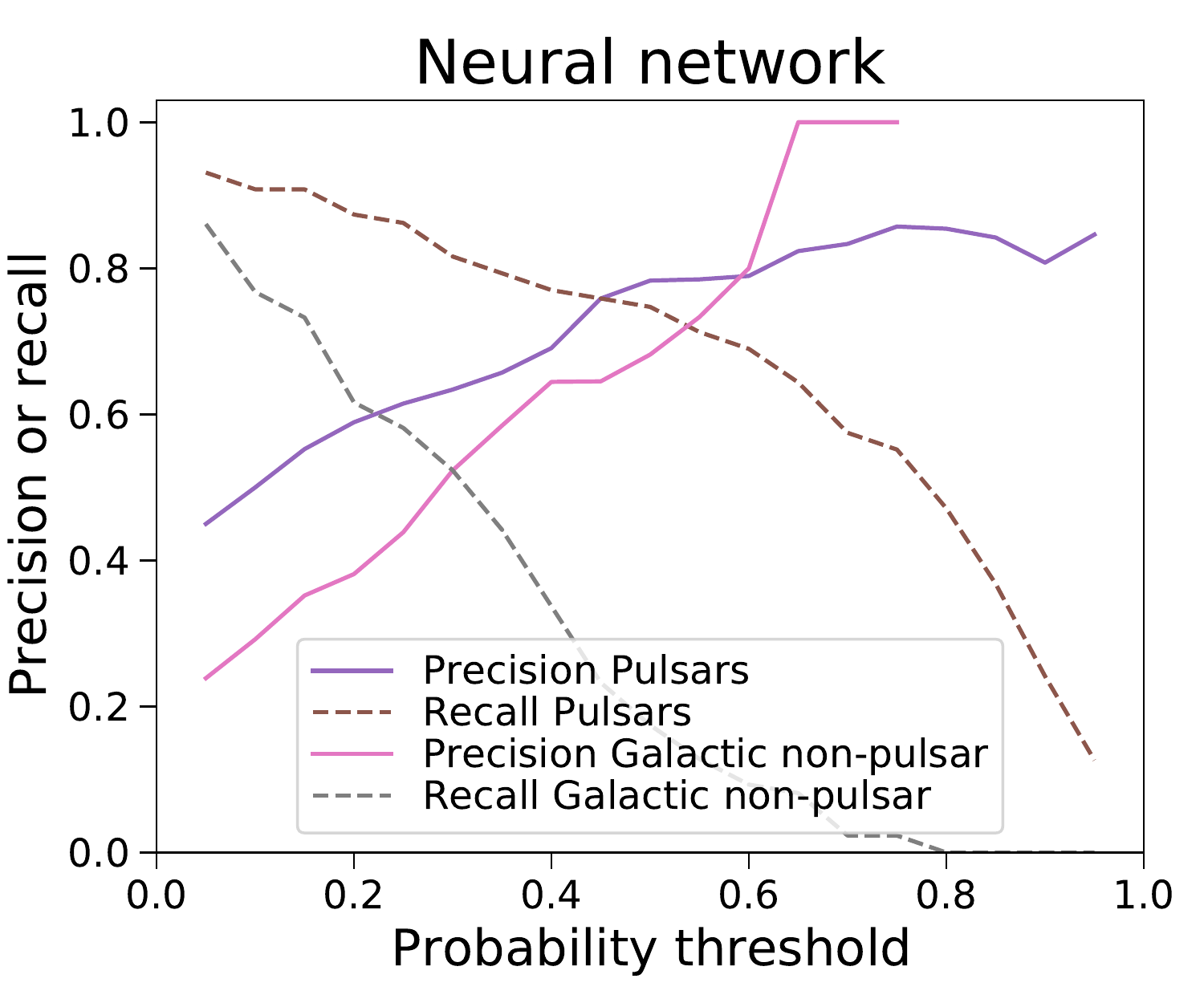}
\caption{Precision and recall for 2- and 3-class classifications as a function of probability threshold.
Left panel: precision and recall for the 2-class classification (in the 3-class curves, the probabilities for the two sub-classes of the Galactic class are summed to determine the probability for the sources to belong to the Galactic class).
The values on the x-axis show the probability above which the sources are classified as Galactic.
Right panel: precision and recall for the two sub-classes of the Galactic class (pulsars and non-pulsar Galactic sources).
The x-axis on this panel shows the probability above which the sources are classified as respectively pulsars and non-pulsar Galactic sources.
}
  \label{fig:prec_recall}
\end{figure}

As the next consistency check we calculate the precision and recall for the Galactic class classification. 
The precision is calculated as the fraction of associated sources, e.g., Galactic, which are also predicted to belong to this class, among all sources predicted to be in this class,
while the recall is calculated as the fraction of associated sources predicted to be in the class among all true (associated) sources in this class:
\begin{equation}
{\rm precision} = \frac{\rm \#\, true\ \&\ predicted}{\rm \#\, predicted};\qquad\qquad
{\rm recall} = \frac{\rm \#\, true\ \&\ predicted}{\rm \#\, true}.
\end{equation}
Precision (recall) is connected to the purity (completeness) of a catalog in astronomical observations.

Left panel in \Cref{fig:prec_recall} shows the precision and recall for the 2-class classification.
We compare the direct training of RF with the 2 classes
to training with 3 classes, where we sum the probabilities of the Galactic class sub-classes in order to obtain the Galactic class probabilities.
Similarly to the ROC curves, there is very little difference between training with 2 and 3 classes for the 2-class precision and recall.
We show precision and recall for the two sub-classes of the Galactic class on the right panel of Figure \ref{fig:prec_recall}.
For example, at the probability threshold of 0.5 the precision  for pulsars and Galactic non-pulsar sources
is 0.85 and 0.73 respectively.
The recall for pulsars and Galactic non-pulsar sources at the probability threshold of 0.5 is respectively 0.71 and 0.44.
This can be compared with the precision and recall for the Galactic class in the 2-class classification (training with 2 classes) of 0.88 and 0.77 respectively.
We find that the fraction of recovered pulsars in the 3-class classification (71\%) is slightly lower than the fraction of all Galactic sources recovered in the 2-class classification (77\%), while the fraction of associated pulsars among all sources predicted to be pulsars in the 3-class case (85\%) is comparable to the fraction of associated Galactic sources among all sources predicted to be Galactic in the 2-class case (88\%).
In addition to that, about half (44\%) of Galactic non-pulsar sources can be recovered in the 3-class classification, while the fraction of associated Galactic non-pulsar sources among the predicted ones is 73\%.
Overall, we find that the 3-class classification provides additional information about the sub-classes of the Galactic class without reduction in precision and recall of the parent class.

\subsection{Reliability diagrams}

\begin{figure}
  \centering
  \includegraphics[width=0.45\textwidth]{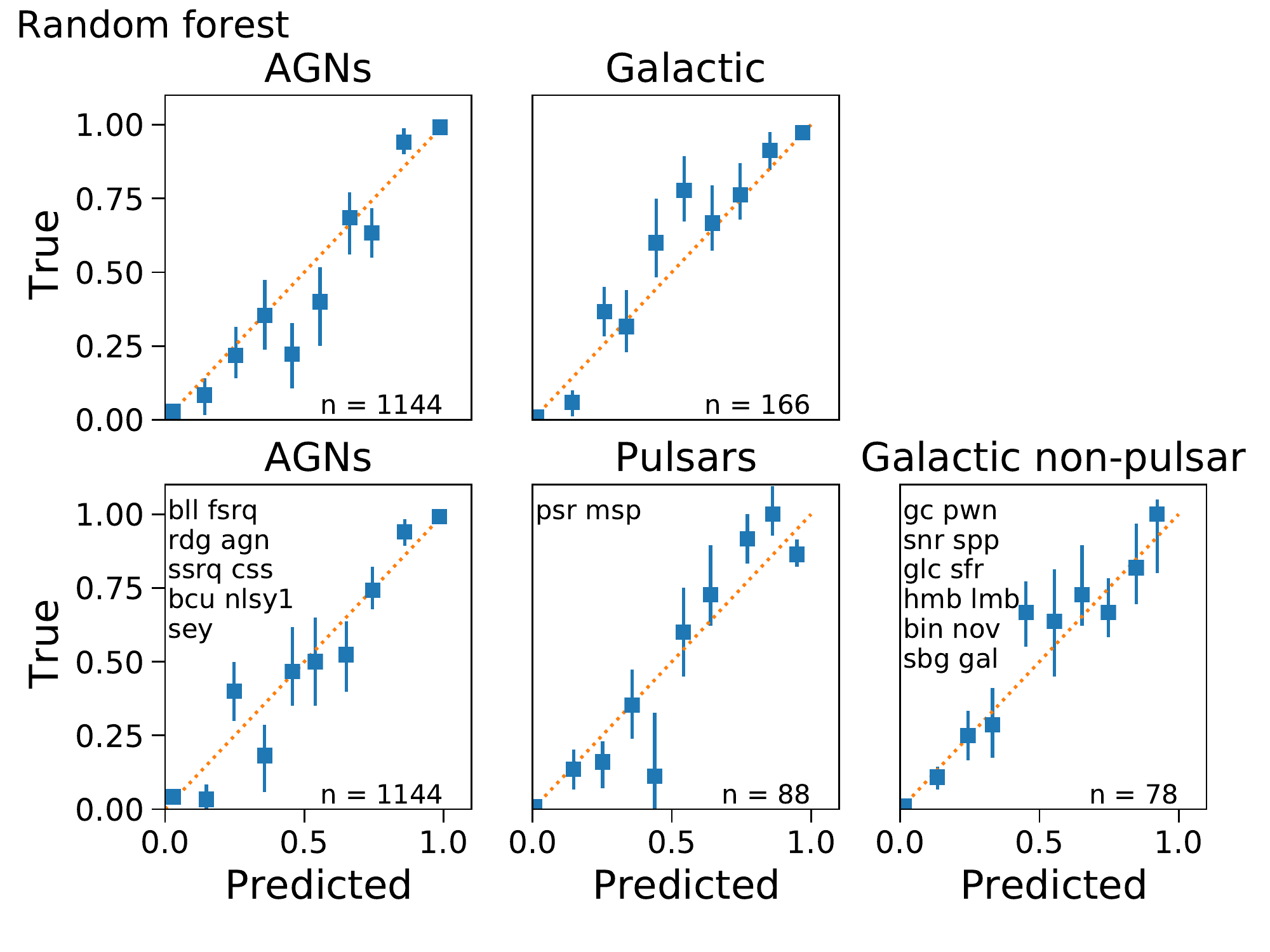}
    \includegraphics[width=0.45\textwidth]{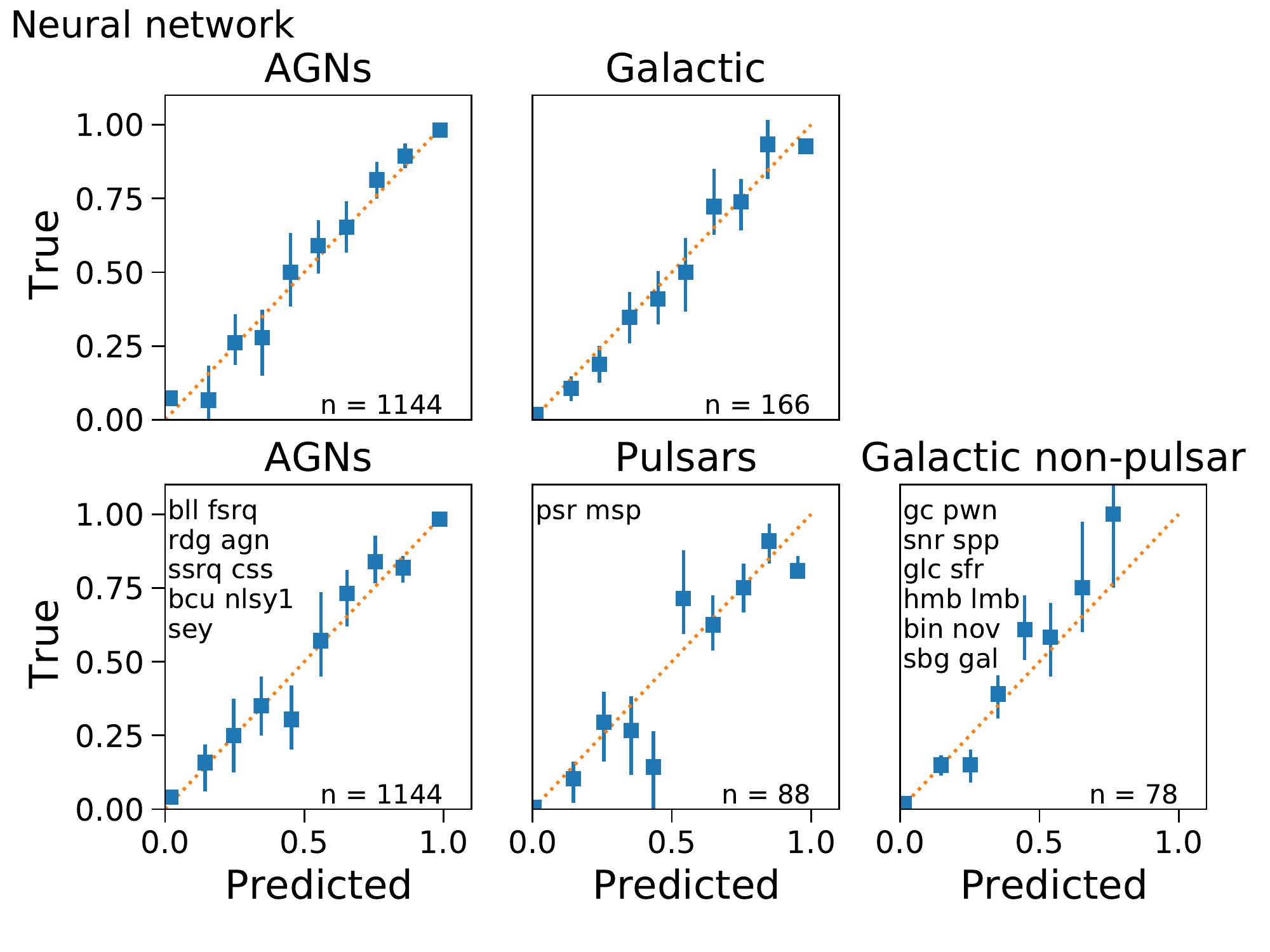}
  \caption{Reliability diagrams for the 2- and 3-class classification for RF (left) and NN (righ) algorithms.
  We show the 4FGL-DR3 labels of the classes in the upper left corners in the 3-class classification (bottom panels).
  The total numbers of sources in the test samples are shown in the lower right corners.
  The red dotted lines show perfect calibration: y = x.
}
  \label{fig:reliability}
\end{figure}

The last check that we performed is the calculation of reliability (or calibration) diagrams for all classes in the 2- and 3-class classification cases.
The top panels in \Cref{fig:reliability} show the reliability diagrams for AGNs and Galactic classes in the 2-class case, while the bottom panels of the figure shows the reliability diagrams for AGNs, pulsars, and Galactic non-pulsar classes in the 3-class case.
The x-axis shows bins in predicted probability: the output score of the RF (left panels) and NN (right panels) algorithms. The x values of the points are computed as the average predicted probability in the bin. The y values are computed as the number of the associated sources in the corresponding class divided by the total number of sources in the probability bin.
The error bars are estimated using 68\% containment of the binomial distribution with the total number of sources given by the number of sources in the probability bin and the probability equal to the center of the bin (divided by the total number of sources in the bin in order to obtain values between 0 and 1).

Technically the output score of the ML algorithms does not need to have a probabilistic interpretation and can be corrected using the calibration diagrams in order to be interpreted as a probability for a source to belong to a certain class.
Nevertheless, it is reassuring that in both the 2- and 3-class cases the RF and NN scores are consistent with the probabilistic interpretation without additional corrections (within the uncertainty bands estimated from the binomial distribution).

\section{Conclusions and Discussion}
\label{sec:conclusions}

In the paper we performed a classification of Fermi-LAT gamma-ray sources into two or three classes. 
Each of these classes encompasses several physical classes reported in the 4FGL-DR3 Fermi-LAT catalog \cite{2022arXiv220111184F}.
We use RF and NN methods for the classification of the sources.
In the calculations we use the class probabilities estimated by the algorithms (rather than the class predictions).
The three classes have an hierarchical structure: two of the three classes are obtained by sub-dividing one of the classes
in the 2-class case, while the third class is the same as in the 2-class case.
One of the main motivations of the work was to study how such subdivision of classes affects the 2-class classification performance 
(in the 3-class case the probabilities of the classes obtained by the subdivision of one of the classes in the 2-class case are summed in order to obtain the probability of the parent class).

We find that, at least in the case of 2- and 3-class classification of Fermi-LAT gamma-ray sources with the RF and NN algorithms, the performance measured by the ROC curves, precision, and recall of the 2-class classification is not degraded by subdivision of one of the classes.
The 3-class classification provides more information about the possible types of sources than the 2-class case, which is valuable for probabilistic classification of unassociated sources.
Additionally, we determine the reliability diagrams and show that both in 2- and 3-class cases the output score of the RF and NN algorithms is consistent with the interpretation as a class probability.
Consequently, the probabilities estimated by the NN and RF algorithms reflect the true fraction of the sources belonging to each class, including low probabilities.
Thus, one can use the sum of probabilities as an estimate of the number of sources in a class in population studies, which include both associated and unassociated sources.
%Since the 2-class classification performance is not degraded with the 3 classes, the 3-class classification has a clear advantage over the 2-class classification, as it provides more information about the classes of sources.

The results obtained in this work show that further subdivision of classes may be possible without degrading performance of classification of the parent classes.
In this paper we discuss an example of the physically motivated subdivision of classes by dividing Galactic sources into pulsars and other Galactic sources.
In general, there are many ways that the 23 physical classes of sources can be separated into 2, 3, 4 or more groups of sources.
If the physical classes were well separated in the feature space, then it would not be necessary to combine classes into larger groups, 
as one could perform the classification with the 23 physical classes.
Provided that there is a different degree of overlap among the classes and that some classes have few members, 
a physically motivated determination of groups of classes may not be optimal from the point of view of classification.
In general, one can also use either unsupervised (e.g., Gaussian mixture model) or supervised (e.g., RF or NN) ML algorithms in order to determine
separation of physical classes into a few larger groups, which provide an optimal classification.
Such determination of optimal groups of physical classes will be the subject of a follow up analysis.

The question of effects of subdivision of classes on classification performance is not limited to the classification of gamma-ray sources, 
but may have important applications in any classification problem where an hierarchical definition of classes is possible.
%For instance, if we classify cats and dogs, subdivision of dogs into large dogs and small dogs or including the breeds of cats and dogs may give an improvement in the overall classification of cats and dogs.
In the example case of the gamma-ray sources considered in this paper, the subdivision of a class does not degrade the performance of the classification
of the parent class, while it provides additional information about the subclasses. 
Thus, in view of application to other classification problems and for generalizations to multi-class classification, 
the question of effects of subdivision of classes on classification performance deserves a further, more detailed study.

\printbibliography

\end{document}